\documentclass[aps,prl,preprint,groupedaddress]{revtex4}
\usepackage[dvips]{graphicx}
\begin{document}
\preprint{}
\title{Tunable Spin Filtering through an Aluminum Nanoparticle}

\author{F. T. Birk, C. E. Malec, D. Davidovi\'c }
\affiliation{School of Physics, Georgia Institute of Technology,
Atlanta, GA 30332}
\date{\today}
\begin{abstract}
Spin-polarized current through an Al nanoparticle in tunnel
contact with two ferromagnets is measured as a function of the
direction of the applied magnetic field. The nanoparticle filters
the spin of injected electrons along a direction specified by the
magnetic field. The characteristic field scale for the filtering
corresponds to a lower limit of $8ns$ for the spin-dephasing time.
Spin polarized current versus applied voltage increases stepwise,
confirming that the spin relaxation time is long only in the
ground state and the low-lying excited states of the nanoparticle.
\end{abstract}

\maketitle

To study and manipulate the properties of the electron spin in a
quantum dot and other electronic structures remains a challenge.
The injection, detection, and coherent manipulation of the
electron spin in GaAs quantum dots have been reported
recently.~\cite{petta3,koppens}  At room temperature, spin
injection, detection, and precession have been measured in
mesoscopic Al strips coupled to ferromagnets by tunnel
contacts.~\cite{jedema} The dephasing time of the precession is
comparable to the electron spin-orbit scattering time
$\tau_{SO}\sim 10^{-10}$s.

In metallic nanoparticles, quantum dot behavior persists at higher
temperatures than in semiconducting quantum dots, making it
relevant to explore spin-relaxation and dephasing in this system.
We expect much longer spin dephasing times in Al nanoparticles
compared to mesoscopic strips  because of the finite size effect.
When the sample size is reduced, both $1/\tau_{SO}$ and the
electron-in-a-box level spacing $\delta$ increase, but $\delta$
increases much faster than $1/\tau_{SO}$.~\cite{wei1} If the
nanoparticles diameter is $D<10nm$, then $\delta> \hbar/\tau_{SO}
$,~\cite{wei1} and the effects of spin-orbit scattering are
suppressed.~\cite{salinas,brouwer,matveev} Evidence of long
spin-relaxation time in metallic nanoparticles have been
reported.~\cite{bernard,wei,mitani1} As will be shown, we can set
a lower limit for the dephasing time in our sample of 8 ns.

Here we report measurements of spin injection, detection, and
basic spin manipulation performed at 4.2 K in an Al nanoparticle
coupled to two ferromagnets by tunnel contacts. The nanoparticle
serves as a quantum dot. The injection and detection of spins is
provided by two ferromagnets, while the manipulation is to filter,
e.g. to project, the spin of injected electrons along the
direction specified by the applied magnetic field. The idea to use
metallic nanoparticle states as spin-filters to study the
magnetization in the ferromagnetic leads was introduced in
Ref.~\cite{deshmukh1} Here we are interested in that filtering to
study the properties of the spin in the nanoparticle, such as
dephasing.  Spin filtering using ferromagnetic tunnel barriers
have been studied recently,~\cite{santos,gajek,ramos} to inject
spins into normal metals without ferromagnetic source and drain
leads. As will be shown, in metallic nanoparticles, the magnetic
field scale for spin filtering is very small, providing tunable
control of the spin direction.

To understand how a magnetic field $\vec B$ influences the
spin-polarized current through the nanoparticle, we consider
electron transport via Zeeman split electron-in-a-box energy
levels. In the ferromagnets, electrons move in an exchange field
oriented along the z-axis, so
the spinors are $|\uparrow>$ and $|\downarrow>$.
In the nanoparticle, the exchange field is negligible, because the
tunnel barriers in our samples are assumed to be very thick, and
the Zeeman splitting in $\vec B$ defines
 spinors
$|\uparrow'>$ and $|\downarrow'>$.

\begin{figure}
\includegraphics[width=0.9\textwidth]{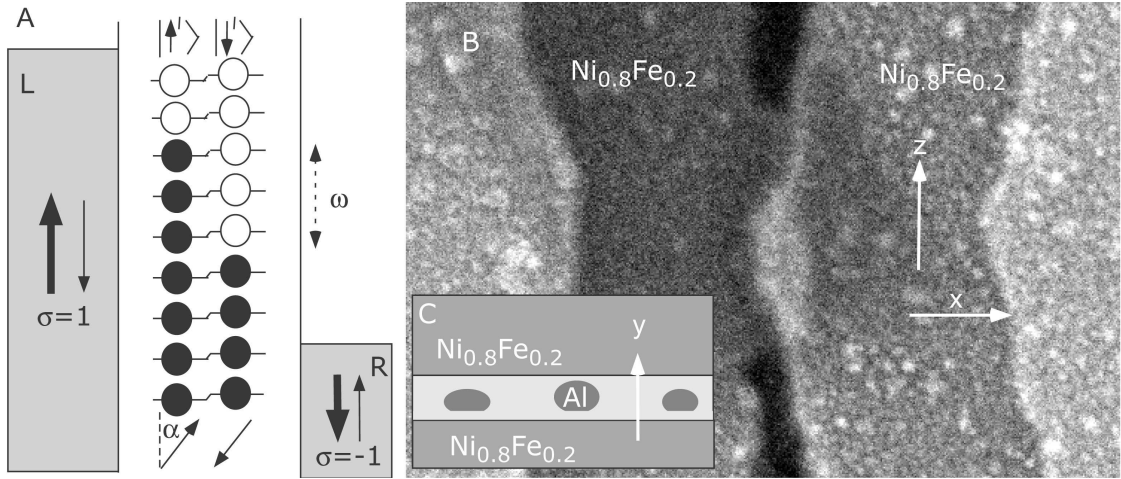}
\caption{A: An excited state of the nanoparticle generated at
finite bias voltage, fully relaxed with respect to spin-conserving
transitions. Filled circles indicate occupied electron-in-a-box
states. The spin expectation value is tilted by the magnetic
field. B: Electron microscope image of a typical device. C: Sketch
of the tunnel junction cross-section. \label{device}}
\end{figure}

We obtain the tunnel-rate $\Gamma_{|\uparrow'>}^r(\alpha)$ between
lead $r$ ($r=R,L$) and an electron-in-a-box level with spin
$|\uparrow'>$ as a function of angle $\alpha$ between $\vec B$ and
the z-axis, as sketched in Fig.~\ref{device}. The magnetization
direction is indicated by parameter $\sigma = \pm 1$ for up and
down directions, respectively. We use a spinor transformation
$|\uparrow'>=cos(\alpha/2)|\uparrow>+sin(\alpha/2)|\downarrow>$,
and consider continuity of the wavefunction at the ferromagnet
interface. With that boundary condition,  the tunnel rate between
$|\uparrow'>$ and the $|\uparrow>$-band in the ferromagnets is
reduced proportionally by a factor of ~\cite{slonczewski}
$cos^2(\alpha/2)$. That tunnel rate can be expressed as
$\Gamma^r(1+\sigma P)cos^2(\alpha/2)$, where $\Gamma^r$ is the
bare tunnel rate defined in Ref.~\cite{wei1} and $P$ is the
spin-polarization in the tunnel density of states in the leads.

If $\alpha\neq 0$, there is also a nonzero transmission between
$|\uparrow'>$ and the spin-down band. Following a similar analysis
as that above, the tunnel rate between $|\uparrow'>$ and the
spin-down band is $\Gamma^r(1-\sigma P)sin^2(\alpha/2)$. The total
tunnel rate between $|\uparrow'>$ and lead $r$ is obtained by
summing over the spin-bands:
$\Gamma_{|\uparrow'>}^r(\alpha)=\Gamma^r[1+\sigma Pcos(\alpha)]$.
Similarly, the total tunnel rate between $|\downarrow'>$ and lead
$r$ is $\Gamma_{|\downarrow'>}^r(\alpha)=\Gamma^r[1-\sigma
Pcos(\alpha)]$.

Overall, $\vec B$ effectively changes the spin-polarization in the
leads from $P$ to $Pcos(\alpha)$. One can obtain the $\vec
B$-dependence of the current using models of spin-polarized
current through the nanoparticle in zero field, by substituting
$P$ with $Pcos(\alpha)$.

The spin-polarized current through the nanoparticle is mediated by
spin accumulation.~\cite{brataas1,weymann2,braun2} Spin
accumulation occurs when the nanoparticle internally excited
states with one spin direction, generated by electron tunnelling
and sketched in Fig.~\ref{device}-A, have higher probability
compared to the states with reversed spin direction.

In the parallel magnetic configuration,
 $\Gamma_{|\uparrow'>}^L(\alpha)/\Gamma_{|\uparrow'>}^R(\alpha)=\Gamma_{|\downarrow'>}^L(\alpha)
 /\Gamma_{|\downarrow'>}^R(\alpha)$,
and sequential electron tunnelling through the nanoparticle will
not cause spin accumulation. In that case $I_{\uparrow\uparrow}$
versus $\vec B$ is constant. This should be contrasted with the
Hanle effect in mesoscopic spin valves,~\cite{johnson} where
$I_{\uparrow\uparrow}$ versus $B_{\perp}$ exhibits a maximum at
$B_{\perp} =0$.

In the antiparallel magnetic state, where $\sigma=1$ and $-1$ for
leads $L$ and $R$, respectively,
$\Gamma_{|\uparrow'>}^L(\alpha)/\Gamma_{|\uparrow'>}^R(\alpha)>\Gamma_{|\downarrow'>}^L(\alpha)
 /\Gamma_{|\downarrow'>}^R(\alpha)$. In that case,  sequential electron tunnelling through the nanoparticle
will cause spin accumulation. Using the spin-accumulation model in
our prior work~\cite{wei1} and substituting $P$ with
$Pcos(\alpha)$, we obtain
\begin{equation}
\Delta I(\alpha)=\Delta I(0)cos^2(\alpha)=\Delta
I(0)\frac{B_z^2}{B_\perp^2+B_z^2} \label{angular1}
\end{equation}
where $\Delta
I(\alpha)=I_{\uparrow\uparrow}-I_{\uparrow\downarrow}(\alpha)$ is
the difference in the current between the parallel and the
antiparallel magnetic configuration. This equation will be
referred to as the spin-filter model.

Our typical device is shown in Fig.~\ref{device}-B. An $Al_2O_3$
tunnel junction is sandwiched between two $Ni80Fe20$ leads in the
overlap region in the center.  Al nanoparticles are embedded in
the tunnel junction, as sketched in Fig.~\ref{device}-C. Many
devices are made simultaneously with variable size of the overlap
region, which varies from a large value to zero. We select the
devices at the threshold of electric conduction, which
significantly enhances the chance that the current between the
leads flows via a single nanoparticle.~\cite{wei} The arrangement
of the leads favors antiparallel magnetic directions. The details
of the fabrication are explained in Ref.~\cite{wei}

At 4.2K, the samples exhibit Coulomb-Blockade (CB). Among the
selected samples, typical parameters are $D\approx 5nm$,
$\delta\sim 1meV$, and $\Gamma\sim MHz$, where $\Gamma$ is the
tunnel rate between the discrete levels and the leads. The
spin-conserving energy relaxation rate in these conditions is
$\gg\Gamma$. At a bias voltage larger than the CB-threshold, only
those nanoparticle states that are fully relaxed with respect to
spin-conserving transitions have significant probability in the
ensemble of excited states generated by tunnelling. One such
excited state is displayed in Fig.~\ref{device}-A.

\begin{figure}
\includegraphics[width=0.9\textwidth]{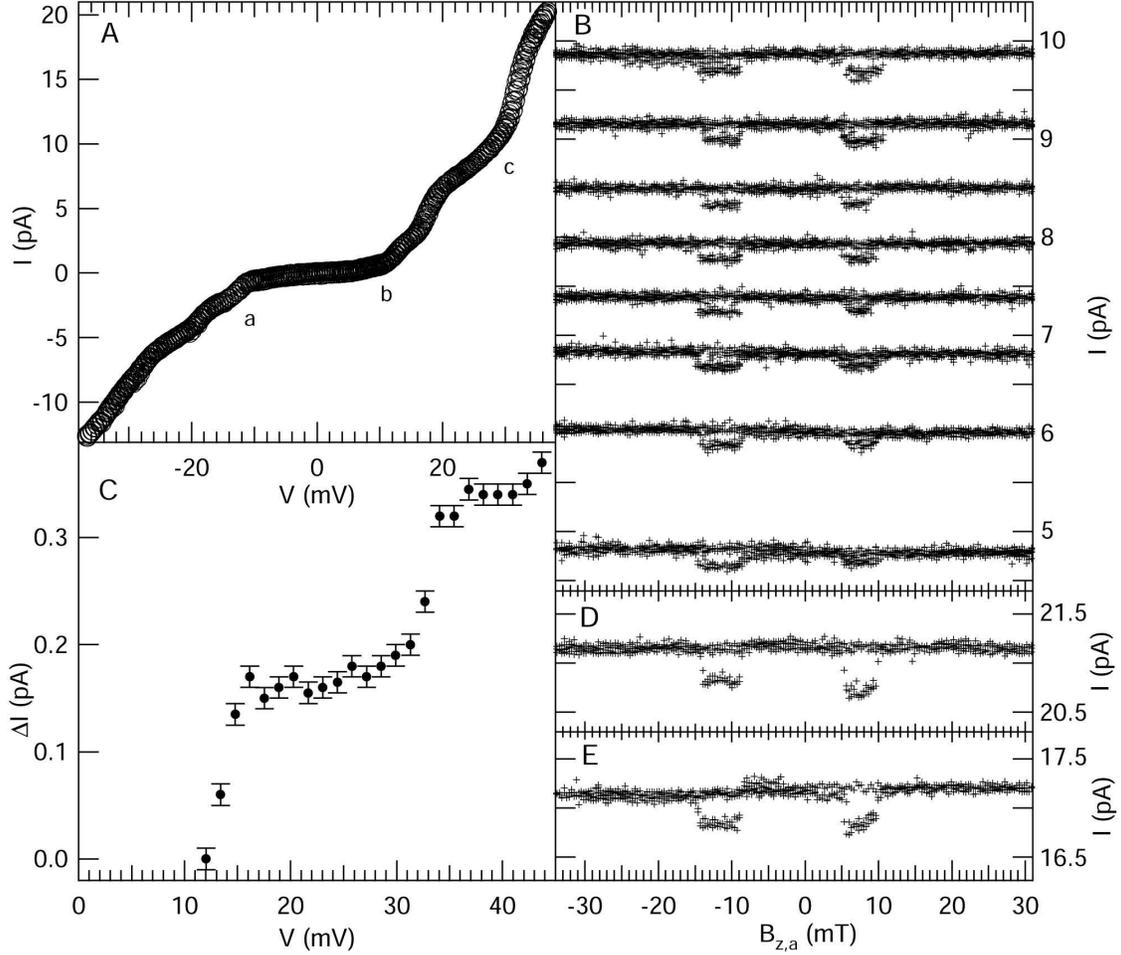}
\caption{A: I-V curve. B: $I$ versus $B_{z,a}$, with increasing
bias voltage. C: $\Delta
I=I_{\uparrow\uparrow}-I_{\uparrow\downarrow}$ versus $V$. D and
E: $I$ versus $B_{z,a}$ at $V=35mV$ and $V=40mV$, respectively.
All data taken at $T=4.2K$.\label{tmr}}
\end{figure}

Fig.~\ref{tmr}-A displays the I-V curve of one device at 4.2K. The
I-V curve exhibits CB. In addition to the conduction thresholds
indicated by the letters a and b, there are thresholds at higher
bias voltages where the slope of the I-V curve increases sharply,
as indicated by the letter c at positive voltage. At the threshold
voltages, additional charged states of the nanoparticle become
energetically available for tunneling, consistent with
CB.~\cite{averin2}\footnote{We compare $dI/dV$ versus $V$ with
theoretical curves~\cite{averin2} to identify the single particle
samples, the criteria introduced in Ref.~\cite{ralph} and used in
our prior work.~\cite{wei} Additionally, the sample is thermally
cycled between 4.2K and 300K several times, each time resulting in
a different I-V curve at 4.2K, because of the different background
charge $q_0$. Various voltage thresholds are traceable with
varying $q_0$, consistent with electron transport in a single
nanoparticle. The threshold c in Fig.~\ref{tmr}-A corresponds
$N+1\to N+2$ tunnel transition, where $N$ is the number of
electrons at $V=0$.}

Current versus parallel applied field $B_{z,a}$ is displayed in
Fig.~\ref{tmr}-B. There are 8 dependencies shown, each obtained at
a different bias voltage. The bias voltage varies by 1.4mV between
successive dependencies. At each bias voltage, the field sweeps
four times in the positive and negative field directions.

The dependencies signal the spin-valve effect. There are two pairs
of magnetic transitions, one for each sweep direction. In a
magnetic transition, the magnetic configuration switches between
parallel and antiparallel, resulting in the current change $\Delta
I=I_{\uparrow\uparrow}-I_{\uparrow\downarrow}$.

In Fig.~\ref{tmr}-B, $\Delta I\approx 0.17pA$ is evidently
independent of $V$. $\Delta I$ versus $V$ is shown explicitly in
Fig.~\ref{tmr}-C. From $V=0$ to $V=32mV$, $\Delta I$ clearly
displays saturation. We also measure the spin-valve signals at
negative bias voltage, and find the same behavior in $\Delta I$
versus $V$, with the same magnitude of $\Delta I$.

The saturation of $\Delta I$ versus $V$ has been reported in our
previous work.~\cite{wei,wei1} $\Delta I$ versus $V$ typically
saturates within the first two or three discrete energy levels
available for tunnelling above the CB-threshold. The saturation
was explained by a rapid decrease in $T_1$ versus energy
difference $\omega$ in a spin-flip transition. To summarize, at
low bias voltage, where $1/T_1(\omega)<\Gamma$ for any excitation
energy $\omega$, $\Delta I\approx c I_{\uparrow\uparrow}$, where
$c$ is on the order of $2P^2$ (the Julliere's value). As the
voltage increases, the range of $\omega$ increases and $\Delta I$
saturates roughly when there is an $\omega$ for which
$1/T_1(\omega)\sim\Gamma$. Self-consistent calculation of the
saturation parameters can be done using a model in
Ref.~\cite{wei1}, which have lead
 to an estimate
$T_1(\delta)\approx 1\mu s$.

Unexpectedly, $\Delta I$ versus $V$ in Fig.~\ref{tmr}-C exhibits
another increase at $V\approx 32mV$. The dependencies $I$ versus
$B_z$ at $V>32mV$ are shown in Fig.~\ref{tmr}-D and E. $\Delta I$
in these figures is approximately two times larger than $\Delta I$
in Fig.~\ref{tmr}-B. To our knowledge, the data in
Fig.~\ref{tmr}B-E is the first observation of a stepwise increase
of the spin-polarized current  with bias voltage through a quantum
dot.

The increase in $\Delta I$ occurs at the conduction threshold
voltage c in Fig.~\ref{tmr}-A. At that voltage, an additional
charged state becomes energetically available for electron
tunneling and starts to contribute significantly to electron
transport. So, the addition of a charged state effectively adds a
channel for spin-accumulation and spin-polarized transport.

At the threshold where the additional charged state becomes
energetically available, there is insufficient energy to generate
internally excited states in the nanoparticle in that additional
charged state. In that case the nanoparticle in the additional
charged state must be in the ground state. If the voltage is
slightly larger than the threshold voltage, then the internal
excitation energy $\omega$ in the additional charged state will be
small and the condition $1/T_1(\omega)<\Gamma$ will be satisfied
again, leading to spin-accumulation. The observation of the
step-wise increase in $\Delta I$ demonstrates the correctness of
our interpretation of the saturation effect in terms of
$T_1(\omega)$ dependence, strengthening our case that we measured
the spin-relaxation time in prior work.~\cite{wei,wei1}

\begin{figure}
\includegraphics[width=0.9\textwidth]{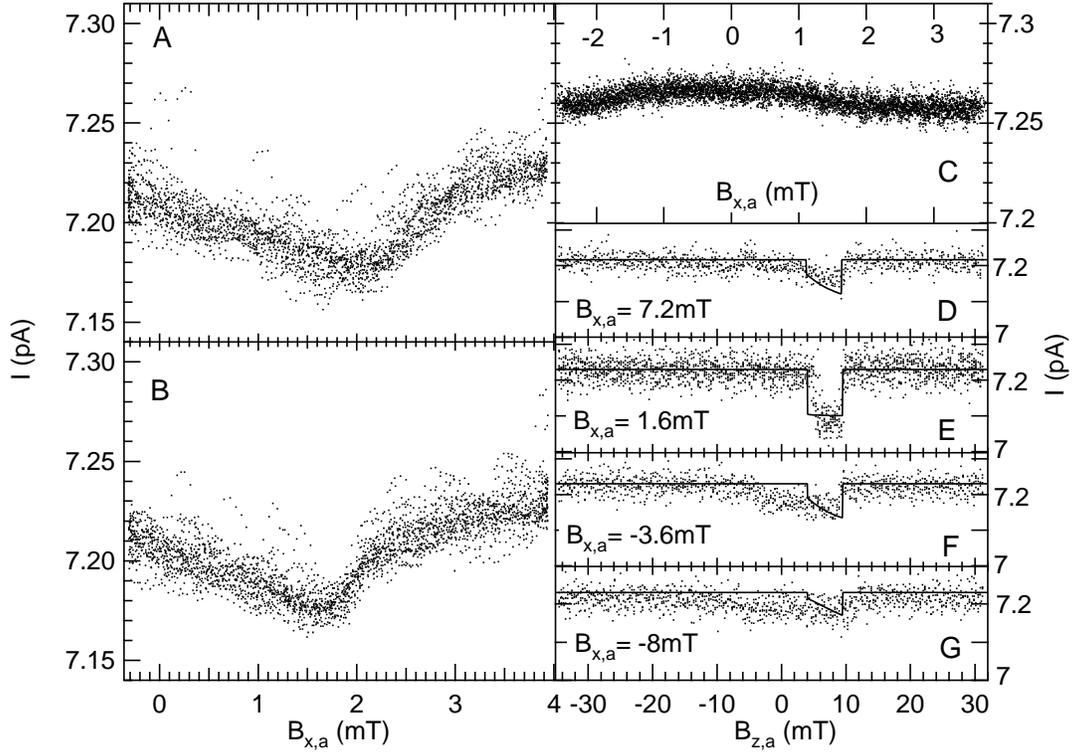}
\caption{A and B: $I_{\uparrow\downarrow}$ versus $B_{x,a}$, for
increasing and decreasing $B_{x,a}$, respectively, at
$B_{y,a}=B_{z,a}=0$. C: $I_{\uparrow\uparrow}$ versus $B_{x,a}$,
at $B_{y,a}=B_{z,a}=0$. D-G: Suppression of the spin-valve signal
with perpendicular applied field. T=4.2K in all
figures.\label{hanle}}
\end{figure}

Now we investigate how a perpendicular field influences
spin-polarized current through the nanoparticle. The
magnetizations are set into the antiparallel configuration using
the spin-valve signal and then the applied field is reduced to
zero. Figs.~\ref{hanle}-A and B display current versus magnetic
field $B_{x,a}$ applied along the $x$-axis, when the magnetic
field is sweeping up and down, respectively. The $x$-axis is
indicated in Fig.~\ref{device}-B. The components of the applied
magnetic field along the easy and the hard axes are zero.


We have carefully measured current versus $B_{x,a}$ in the
parallel magnetic configuration, when $B_{z,a}=B_{y,a}=0$. The
results are shown in Fig.~\ref{hanle}-C. Comparing
Figs.~\ref{hanle}-A, B, and C, it is concluded that $I$ versus
$B_{x,a}$ exhibits a minimum in the antiparallel magnetic
configuration  and $I$ versus $B_{x,a}$ is constant in the
parallel magnetic configuration.

The minimum center is offset and the amplitude of the minimum,
$\approx 0.05pA$, is smaller than $\Delta I=0.17pA$ measured in
the spin-valve signal in Fig.~\ref{tmr}. Comparing
Figs.~\ref{hanle}-A and B, we conclude that the minimum is
reversible with magnetic field, although there is a weak
hysteresis, of approximately $0.2mT$.

The spin-valve signal is suppressed with the applied perpendicular
field, as shown in Figs.~\ref{hanle}-D, E, F, and G. The strongest
spin-valve signal, shown in Fig.~\ref{hanle}-E, is measured around
the perpendicular field at the minimum of Figs.~\ref{hanle}-A and
B. In a strong perpendicular field, Fig.~\ref{hanle}-D,F, and G,
the magnetic transitions from parallel into antiparallel magnetic
state become significantly weakened; as $B_{z,a}$ approaches the
magnetic transition from antiparallel into the parallel magnetic
state, there is now a gradual decrease in current. The transitions
from antiparallel to parallel magnetic state in a strong
perpendicular field remain resolved, but they are weakened
proportionally to the magnitude of the perpendicular field. The
characteristic perpendicular field that weakens the spin-valve
signal is much larger than the width of the minimum in
Fig.~\ref{hanle}-A and B.

The measurements are in agreement with the spin-filter model
described in the introduction. We must rule out a possibility that
the dependencies in Figs.~\ref{hanle} are caused by rotation of
the magnetizations in response to the perpendicular field as it
could explain the data so far presented. Starting from the
antiparallel magnetic configuration, a rotation would vary the
magnetizations from antiparallel to parallel, leading to a minimum
in $I$ versus $B_{x,a}$. Starting from the parallel magnetic
configuration, that rotation may not vary the angle between the
magnetizations, thus maintaining the leads in a parallel state,
and resulting in no dependence with perpendicular field.

\begin{figure}
\includegraphics[width=0.9\textwidth]{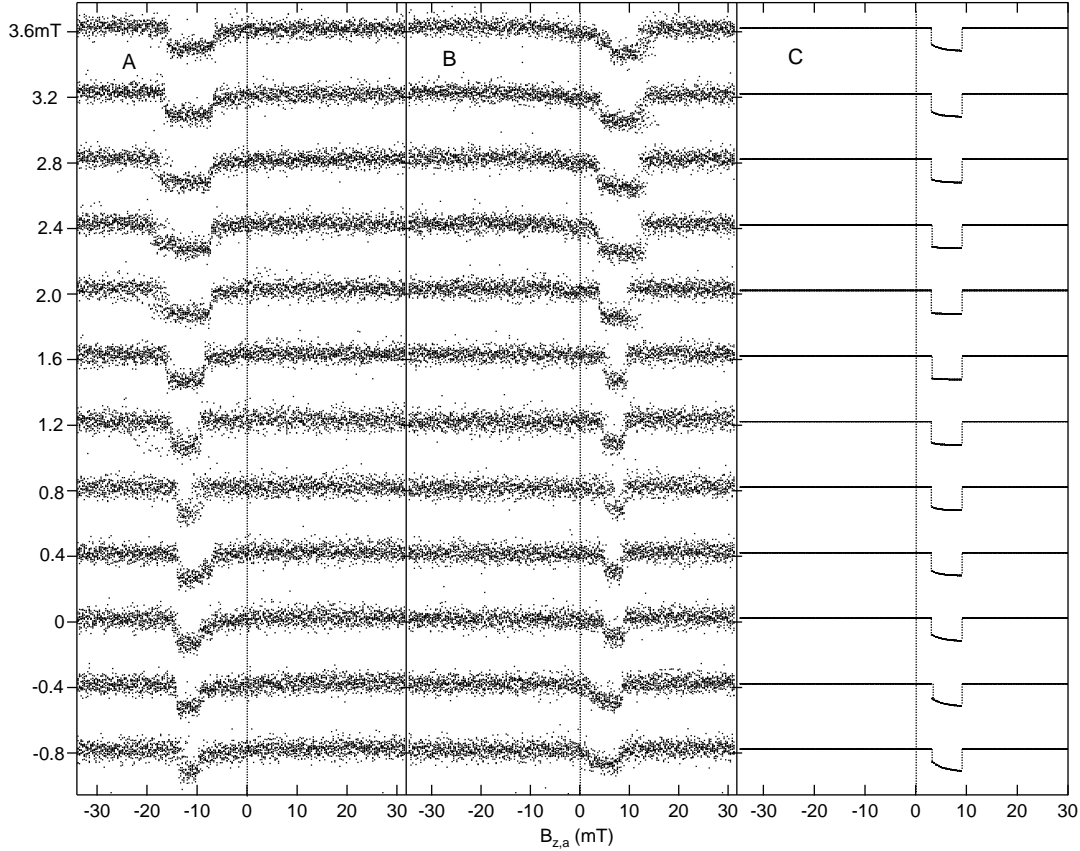}
\caption{A and B: Spin-valve signals versus weak perpendicular
field $B_{x,a}$, for decreasing and increasing $B_{z,a}$,
respectively, at $T=4.2K$. C: Calculated spin-valve signal
corresponding to Fig.~\ref{tmr1}-B.\label{tmr1}}
\end{figure}

Fig.~\ref{tmr1} displays a family of spin-valve signals measured
at different perpendicular fields $B_{x,a}$, which vary within the
field range of the minimum in Fig.~\ref{hanle}-A and B. $B_{x,a}$
is indicated on the vertical axis. To trace the spin-valve signal
versus $B_{x,a}$, successive $I$ versus $B_{z,a}$ curves are
offset by $0.36pA$. The spin-valve signal in Fig.~\ref{tmr1} is
weakly affected by the perpendicular field, which demonstrates
that the parallel and the antiparallel magnetic configurations are
stable in the range where the minimum in Fig.~\ref{hanle}-A and B
is observed.

We analyze the dependencies in Figs.~\ref{hanle} and ~\ref{tmr1}
using Eq.~\ref{angular1}, where $\vec B=\vec B_a+\vec B_{l}$. The
fields $\vec B_{a}$ and $\vec B_{l}$ are the applied and the local
field, respectively. $\vec B_{l}$ arises in part from the
demagnetizing field generated by the leads. $\Delta I(0)$ in
Eq.~\ref{angular1} is obtained as the maximum value of
$I_{\uparrow\uparrow}-I_{\uparrow\downarrow}$ as a function of
perpendicular field in Fig.~\ref{tmr1}

The amplitude, the full-width-half-minimum, and the center of
curves in Fig.~\ref{hanle}A and B should correspond to $\Delta
I\frac{B_{z,l}^2}{B_{z,l}^2+B_{y,l}^2}$,
$2\sqrt{B_{z,l}^2+B_{y,l}^2}$, and $-B_{x,l}$ in
Eq.~\ref{angular1}. That leads to $\vec {B_l}=(-1.8mT,\pm
0.63mT,\pm 0.41mT)$.

Next, using this local field and Eq.~\ref{angular1}, we calculate
the spin-valve signal, using fixed coercive fields in the leads of
$4$ and $9mT$. The results of the calculation are indicated by the
lines in Fig.~\ref{hanle} D-G, showing good agreement. Thus, the
effect of the applied perpendicular field on the spin-valve signal
is well explained by the spin-filter model.

The reason that the spin-valve signal in Fig.~\ref{tmr1} is weakly
affected by the perpendicular field, compared to the effect at
$B_{z,a}=0$ shown in Figs.~\ref{hanle}A and B, is that in the spin
valve signal, $B_{z,a}$ is large compared to $B_{x,a}$ in the
antiparallel magnetic configuration, so the spin-valve signal has
reduced sensitivity to the perpendicular field. Fig.~\ref{tmr1}-C
displays the spin-valve signal versus increasing $B_{x,a}$
calculated from Eq.~\ref{angular1} as explained above, showing
that the perpendicular field in Fig.~\ref{tmr1} is weak to
suppress the amplitude of the spin-valve signal. The calculation
for decreasing $B_{x,a}$ leads to the same conclusion.

Initially, we studied the effect of the magnetic field applied
along the hard axis ($B_{y,a}$) in this sample. The spin valve
signal was significantly weakened with a strong hard axis field,
analogous to the effect in Figs.~\ref{hanle} D,F, and G. But, in
zero applied field, $B_{x,a}=B_{z,a}=0$, we could not resolve any
dependence in $I$ versus $B_{y,a}$ beyond noise. The absence of
minimum with $B_{y,a}$ is explained by the large component of the
local field along the $x$ direction. Using Eq.~\ref{angular1}, the
amplitude of the minimum in $I$ versus $B_{y,a}$ should be $\Delta
I\frac{B_{z,l}^2}{B_{z,l}^2+B_{x,l}^2}\approx 8fA$, which is less
than the noise.

Since the dominant component of $\vec {B_l}$ is along $x$, this
suggests that the local field is generated by a domain magnetized
along the x-direction, in the vicinity of the nanoparticle. Such a
domain would explain the hysteresis and asymmetry in
Figs.~\ref{hanle}-A and B, if the domain wall moved in response to
changing $B_{x,a}$. In that case the local field would not be
completely independent of the applied field. Hysteresis of the
domain wall motion would lead to a hysteresis in $B_{x,l}$. The
asymmetry of the minimum could also be attributed to the
dependence of $B_{x,l}$ on $B_{x,a}$. But the hysteresis in
$B_{x,l}$ is only $0.2mT$, which is about 10\%. So, in the lowest
order of approximation, it can be assumed that the local field is
constant in our applied field range.

The spin filter model is valid if the dephasing is weak. In a
single metallic nanoparticle, the dephasing is caused by temporal
fluctuations of the magnetic field, which can randomize the spin
angle with respect to the z-axis. The current through the
nanoparticle in contact with ferromagnetic leads in the presence
of dephasing was calculated recently by Braun et al.~\cite{braun}
Their model leads to the formula
\begin{equation}
\Delta
I(\alpha)\sim\frac{\omega_{B_z}^2}{\nu_S^2+\omega_{B_z}^2+\omega_{B_\perp}^2},
\label{angular2}
\end{equation}
where $\vec \omega_B=g\mu_B \vec B/\hbar$. This dependence is
identical to that in Eq.~\ref{angular1}, except for the dephasing
term in the denominator. The dephasing term increases the width of
the minimum in current versus perpendicular field. The dephasing
is negligible if $g\mu_B B > \nu_S$. Eq.~\ref{angular2} can be
used to obtain a lower limit for the spin-dephasing time in the
nanoparticle:
$T_2=1/\nu_S>\hbar/g\mu_B\sqrt{B_{z,l}^2+B_{y,l}^2}=8nS$. The
dephasing time in the nanoparticle is enhanced compared to that in
mesoscopic Al strips,~\cite{jedema} and it is more in line with
the lower bounds of $T_2$ measured in GaAs quantum
dots.~\cite{petta3}

In conclusion, spin polarized current through an Aluminum
nanoparticle is very sensitive to the direction of the magnetic
field and consistent with a picture in which the nanoparticle
states filter spin polarized current by selecting the spinor
component specified by the magnetic field. A magnetic field
applied perpendicular to the direction of the magnetizations
suppresses spin polarized current. A lower bound of the spin
dephasing time, 8ns, is obtained from the characteristic field for
that suppression. As a function of bias voltage, a stepwise
increase in spin polarized current is observed when an additional
charged state of the nanoparticle becomes conductive, confirming
that spin-relaxation time is $\sim \mu s$ only if the nanoparticle
is in the ground state or very close to it.

This research is supported by the DOE grant DE-FG02-06ER46281
 and David and Lucile Packard Foundation grant
2000-13874.

\bibliography{career1}

\begin{thebibliography}{22}
\expandafter\ifx\csname natexlab\endcsname\relax\def\natexlab#1{#1}\fi
\expandafter\ifx\csname bibnamefont\endcsname\relax
  \def\bibnamefont#1{#1}\fi
\expandafter\ifx\csname bibfnamefont\endcsname\relax
  \def\bibfnamefont#1{#1}\fi
\expandafter\ifx\csname citenamefont\endcsname\relax
  \def\citenamefont#1{#1}\fi
\expandafter\ifx\csname url\endcsname\relax
  \def\url#1{\texttt{#1}}\fi
\expandafter\ifx\csname urlprefix\endcsname\relax\def\urlprefix{URL }\fi
\providecommand{\bibinfo}[2]{#2}
\providecommand{\eprint}[2][]{\url{#2}}

\bibitem[{\citenamefont{Petta et~al.}(2004)\citenamefont{Petta, Johnson,
  Taylor, Laird, A, Lukin, Marcus, Hanson, and Gossard}}]{petta3}
\bibinfo{author}{\bibfnamefont{J.~R.} \bibnamefont{Petta}},
  \bibinfo{author}{\bibfnamefont{A.~C.} \bibnamefont{Johnson}},
  \bibinfo{author}{\bibfnamefont{J.~M.} \bibnamefont{Taylor}},
  \bibinfo{author}{\bibfnamefont{E.~A.} \bibnamefont{Laird}},
  \bibinfo{author}{\bibfnamefont{A.~Y.} \bibnamefont{A}},
  \bibinfo{author}{\bibfnamefont{M.~D.} \bibnamefont{Lukin}},
  \bibinfo{author}{\bibfnamefont{C.~M.} \bibnamefont{Marcus}},
  \bibinfo{author}{\bibfnamefont{M.~P.} \bibnamefont{Hanson}},
  \bibnamefont{and} \bibinfo{author}{\bibfnamefont{A.~C.}
  \bibnamefont{Gossard}}, \bibinfo{journal}{Science}
  \textbf{\bibinfo{volume}{309}}, \bibinfo{pages}{2180} (\bibinfo{year}{2004}).

\bibitem[{\citenamefont{Koppens et~al.}(2006)\citenamefont{Koppens, Buizert,
  Tielrooij, Vink, Nowack, Meunier, Kouwenhoven, and Vandersypen}}]{koppens}
\bibinfo{author}{\bibfnamefont{F.~H.~L.} \bibnamefont{Koppens}},
  \bibinfo{author}{\bibfnamefont{C.}~\bibnamefont{Buizert}},
  \bibinfo{author}{\bibfnamefont{K.~J.} \bibnamefont{Tielrooij}},
  \bibinfo{author}{\bibfnamefont{I.~T.} \bibnamefont{Vink}},
  \bibinfo{author}{\bibfnamefont{K.~C.} \bibnamefont{Nowack}},
  \bibinfo{author}{\bibfnamefont{T.}~\bibnamefont{Meunier}},
  \bibinfo{author}{\bibfnamefont{L.~P.} \bibnamefont{Kouwenhoven}},
  \bibnamefont{and} \bibinfo{author}{\bibfnamefont{L.~M.~K.}
  \bibnamefont{Vandersypen}}, \bibinfo{journal}{Nature}
  \textbf{\bibinfo{volume}{442}}, \bibinfo{pages}{766} (\bibinfo{year}{2006}).

\bibitem[{\citenamefont{Jedema et~al.}(2001)\citenamefont{Jedema, Filip, and
  van Wees}}]{jedema}
\bibinfo{author}{\bibfnamefont{F.~J.} \bibnamefont{Jedema}},
  \bibinfo{author}{\bibfnamefont{A.~T.} \bibnamefont{Filip}}, \bibnamefont{and}
  \bibinfo{author}{\bibfnamefont{B.~J.} \bibnamefont{van Wees}},
  \bibinfo{journal}{Nature} \textbf{\bibinfo{volume}{410}},
  \bibinfo{pages}{345} (\bibinfo{year}{2001}).

\bibitem[{\citenamefont{Wei et~al.}(2008)\citenamefont{Wei, Malec, and
  Davidovic}}]{wei1}
\bibinfo{author}{\bibfnamefont{Y.~G.} \bibnamefont{Wei}},
  \bibinfo{author}{\bibfnamefont{C.~E.} \bibnamefont{Malec}}, \bibnamefont{and}
  \bibinfo{author}{\bibfnamefont{D.}~\bibnamefont{Davidovic}},
  \bibinfo{journal}{Phys. Rev. B} \textbf{\bibinfo{volume}{78}},
  \bibinfo{pages}{035435} (\bibinfo{year}{2008}).

\bibitem[{\citenamefont{Salinas et~al.}(1999)\citenamefont{Salinas, Gueron,
  Ralph, Black, and Tinkham}}]{salinas}
\bibinfo{author}{\bibfnamefont{D.~G.} \bibnamefont{Salinas}},
  \bibinfo{author}{\bibfnamefont{S.}~\bibnamefont{Gueron}},
  \bibinfo{author}{\bibfnamefont{D.~C.} \bibnamefont{Ralph}},
  \bibinfo{author}{\bibfnamefont{C.~T.} \bibnamefont{Black}}, \bibnamefont{and}
  \bibinfo{author}{\bibfnamefont{M.}~\bibnamefont{Tinkham}},
  \bibinfo{journal}{Phys. Rev. B} \textbf{\bibinfo{volume}{60}},
  \bibinfo{pages}{6137} (\bibinfo{year}{1999}).

\bibitem[{\citenamefont{Brouwer et~al.}(2000)\citenamefont{Brouwer, Waintal,
  and Halperin}}]{brouwer}
\bibinfo{author}{\bibfnamefont{P.~W.} \bibnamefont{Brouwer}},
  \bibinfo{author}{\bibfnamefont{X.}~\bibnamefont{Waintal}}, \bibnamefont{and}
  \bibinfo{author}{\bibfnamefont{B.~I.} \bibnamefont{Halperin}},
  \bibinfo{journal}{Phys. Rev. Lett.} \textbf{\bibinfo{volume}{85}},
  \bibinfo{pages}{369} (\bibinfo{year}{2000}).

\bibitem[{\citenamefont{Matveev et~al.}(2000)\citenamefont{Matveev, Glazman,
  and Larkin}}]{matveev}
\bibinfo{author}{\bibfnamefont{K.~A.} \bibnamefont{Matveev}},
  \bibinfo{author}{\bibfnamefont{L.~I.} \bibnamefont{Glazman}},
  \bibnamefont{and} \bibinfo{author}{\bibfnamefont{A.~I.}
  \bibnamefont{Larkin}}, \bibinfo{journal}{Phys. Rev. Lett.}
  \textbf{\bibinfo{volume}{85}}, \bibinfo{pages}{2789} (\bibinfo{year}{2000}).

\bibitem[{\citenamefont{Bernand-Mantel
  et~al.}(2006)\citenamefont{Bernand-Mantel, Seneor, Lidgi, Munoz, Cros, Fusil,
  Bouzehouane, Deranlot, Vaures, Petroff et~al.}}]{bernard}
\bibinfo{author}{\bibfnamefont{A.}~\bibnamefont{Bernand-Mantel}},
  \bibinfo{author}{\bibfnamefont{P.}~\bibnamefont{Seneor}},
  \bibinfo{author}{\bibfnamefont{N.}~\bibnamefont{Lidgi}},
  \bibinfo{author}{\bibfnamefont{M.}~\bibnamefont{Munoz}},
  \bibinfo{author}{\bibfnamefont{V.}~\bibnamefont{Cros}},
  \bibinfo{author}{\bibfnamefont{S.}~\bibnamefont{Fusil}},
  \bibinfo{author}{\bibfnamefont{K.}~\bibnamefont{Bouzehouane}},
  \bibinfo{author}{\bibfnamefont{C.}~\bibnamefont{Deranlot}},
  \bibinfo{author}{\bibfnamefont{A.}~\bibnamefont{Vaures}},
  \bibinfo{author}{\bibfnamefont{F.}~\bibnamefont{Petroff}},
  \bibnamefont{et~al.}, \bibinfo{journal}{Appl. Phys. Lett.}
  \textbf{\bibinfo{volume}{89}}, \bibinfo{pages}{062502}
  (\bibinfo{year}{2006}).

\bibitem[{\citenamefont{Wei et~al.}(2007)\citenamefont{Wei, Malec, and
  Davidovic}}]{wei}
\bibinfo{author}{\bibfnamefont{Y.~G.} \bibnamefont{Wei}},
  \bibinfo{author}{\bibfnamefont{C.~E.} \bibnamefont{Malec}}, \bibnamefont{and}
  \bibinfo{author}{\bibfnamefont{D.}~\bibnamefont{Davidovic}},
  \bibinfo{journal}{Phys. Rev. B} \textbf{\bibinfo{volume}{76}},
  \bibinfo{pages}{195327} (\bibinfo{year}{2007}).

\bibitem[{\citenamefont{Mitani et~al.}(2008)\citenamefont{Mitani, Nogi, Wang,
  Yakushiji, Ernult, and K}}]{mitani1}
\bibinfo{author}{\bibfnamefont{S.}~\bibnamefont{Mitani}},
  \bibinfo{author}{\bibfnamefont{Y.}~\bibnamefont{Nogi}},
  \bibinfo{author}{\bibfnamefont{H.}~\bibnamefont{Wang}},
  \bibinfo{author}{\bibfnamefont{K.}~\bibnamefont{Yakushiji}},
  \bibinfo{author}{\bibfnamefont{F.}~\bibnamefont{Ernult}}, \bibnamefont{and}
  \bibinfo{author}{\bibfnamefont{K.~T.} \bibnamefont{K}},
  \bibinfo{journal}{Appl. Phys. Lett.} \textbf{\bibinfo{volume}{92}},
  \bibinfo{pages}{152509} (\bibinfo{year}{2008}).

\bibitem[{\citenamefont{Deshmukh and Ralph}(2002)}]{deshmukh1}
\bibinfo{author}{\bibfnamefont{M.~M.} \bibnamefont{Deshmukh}} \bibnamefont{and}
  \bibinfo{author}{\bibfnamefont{D.~C.} \bibnamefont{Ralph}},
  \bibinfo{journal}{Phys. Rev. Lett} \textbf{\bibinfo{volume}{89}},
  \bibinfo{pages}{266803} (\bibinfo{year}{2002}).

\bibitem[{\citenamefont{Santos and Moodera}(2004)}]{santos}
\bibinfo{author}{\bibfnamefont{T.~S.} \bibnamefont{Santos}} \bibnamefont{and}
  \bibinfo{author}{\bibfnamefont{J.~S.} \bibnamefont{Moodera}},
  \bibinfo{journal}{Phys. Rev. B} \textbf{\bibinfo{volume}{69}},
  \bibinfo{pages}{241203(R)} (\bibinfo{year}{2004}).

\bibitem[{\citenamefont{Gajek et~al.}(2005)\citenamefont{Gajek, Bibes,
  Barthelemy, Bouzehouane, Fusil, Varela, Fontcuberta, and Fert}}]{gajek}
\bibinfo{author}{\bibfnamefont{M.}~\bibnamefont{Gajek}},
  \bibinfo{author}{\bibfnamefont{M.}~\bibnamefont{Bibes}},
  \bibinfo{author}{\bibfnamefont{A.}~\bibnamefont{Barthelemy}},
  \bibinfo{author}{\bibfnamefont{K.}~\bibnamefont{Bouzehouane}},
  \bibinfo{author}{\bibfnamefont{S.}~\bibnamefont{Fusil}},
  \bibinfo{author}{\bibfnamefont{M.}~\bibnamefont{Varela}},
  \bibinfo{author}{\bibfnamefont{J.}~\bibnamefont{Fontcuberta}},
  \bibnamefont{and} \bibinfo{author}{\bibfnamefont{A.}~\bibnamefont{Fert}},
  \bibinfo{journal}{Phys. Rev. B} \textbf{\bibinfo{volume}{72}},
  \bibinfo{pages}{020406(R)} (\bibinfo{year}{2005}).

\bibitem[{\citenamefont{Ramos et~al.}(2007)\citenamefont{Ramos, Guittet,
  Moussy, Mattana, Deranlot, Petroff, and Gatel}}]{ramos}
\bibinfo{author}{\bibfnamefont{A.~V.} \bibnamefont{Ramos}},
  \bibinfo{author}{\bibfnamefont{M.-J.} \bibnamefont{Guittet}},
  \bibinfo{author}{\bibfnamefont{J.~B.} \bibnamefont{Moussy}},
  \bibinfo{author}{\bibfnamefont{R.}~\bibnamefont{Mattana}},
  \bibinfo{author}{\bibfnamefont{C.}~\bibnamefont{Deranlot}},
  \bibinfo{author}{\bibfnamefont{F.}~\bibnamefont{Petroff}}, \bibnamefont{and}
  \bibinfo{author}{\bibfnamefont{C.}~\bibnamefont{Gatel}},
  \bibinfo{journal}{Appl. Phys. Lett.} \textbf{\bibinfo{volume}{91}},
  \bibinfo{pages}{122107} (\bibinfo{year}{2007}).

\bibitem[{\citenamefont{Slonczewski}(1989)}]{slonczewski}
\bibinfo{author}{\bibfnamefont{J.~C.} \bibnamefont{Slonczewski}},
  \bibinfo{journal}{Phys. Rev. B} \textbf{\bibinfo{volume}{39}},
  \bibinfo{pages}{6995} (\bibinfo{year}{1989}).

\bibitem[{\citenamefont{Brataas et~al.}(1999)\citenamefont{Brataas, Nazarov,
  Inoue, and Bauer}}]{brataas1}
\bibinfo{author}{\bibfnamefont{A.}~\bibnamefont{Brataas}},
  \bibinfo{author}{\bibfnamefont{Y.~V.} \bibnamefont{Nazarov}},
  \bibinfo{author}{\bibfnamefont{J.}~\bibnamefont{Inoue}}, \bibnamefont{and}
  \bibinfo{author}{\bibfnamefont{G.~E.~W.} \bibnamefont{Bauer}},
  \bibinfo{journal}{Phys. Rev. B} \textbf{\bibinfo{volume}{59}},
  \bibinfo{pages}{93} (\bibinfo{year}{1999}).

\bibitem[{\citenamefont{Weymann et~al.}(20053)\citenamefont{Weymann, Konig,
  Martinek, Barnas, and Schon}}]{weymann2}
\bibinfo{author}{\bibfnamefont{I.}~\bibnamefont{Weymann}},
  \bibinfo{author}{\bibfnamefont{J.}~\bibnamefont{Konig}},
  \bibinfo{author}{\bibfnamefont{J.}~\bibnamefont{Martinek}},
  \bibinfo{author}{\bibfnamefont{J.}~\bibnamefont{Barnas}}, \bibnamefont{and}
  \bibinfo{author}{\bibfnamefont{G.}~\bibnamefont{Schon}},
  \bibinfo{journal}{Phys. Rev. B} \textbf{\bibinfo{volume}{72}},
  \bibinfo{pages}{115334} (\bibinfo{year}{20053}).

\bibitem[{\citenamefont{Braun et~al.}(2004)\citenamefont{Braun, Konig, and
  Martinek}}]{braun2}
\bibinfo{author}{\bibfnamefont{M.}~\bibnamefont{Braun}},
  \bibinfo{author}{\bibfnamefont{J.}~\bibnamefont{Konig}}, \bibnamefont{and}
  \bibinfo{author}{\bibfnamefont{J.}~\bibnamefont{Martinek}},
  \bibinfo{journal}{Phys. Rev. B} \textbf{\bibinfo{volume}{70}},
  \bibinfo{pages}{195345} (\bibinfo{year}{2004}).

\bibitem[{\citenamefont{Johnson and Silsbee}(1985)}]{johnson}
\bibinfo{author}{\bibfnamefont{M.}~\bibnamefont{Johnson}} \bibnamefont{and}
  \bibinfo{author}{\bibfnamefont{R.~H.} \bibnamefont{Silsbee}},
  \bibinfo{journal}{Phys. Rev. Lett.} \textbf{\bibinfo{volume}{55}},
  \bibinfo{pages}{1790} (\bibinfo{year}{1985}).

\bibitem[{\citenamefont{Averin and Likharev}(1991)}]{averin2}
\bibinfo{author}{\bibfnamefont{D.~V.} \bibnamefont{Averin}} \bibnamefont{and}
  \bibinfo{author}{\bibfnamefont{K.~K.} \bibnamefont{Likharev}}, in
  \emph{\bibinfo{booktitle}{Mesoscopic Phenomena in Solids}}, edited by
  \bibinfo{editor}{\bibfnamefont{B.~L.} \bibnamefont{Altshuler}},
  \bibinfo{editor}{\bibfnamefont{P.~L.} \bibnamefont{Lee}}, \bibnamefont{and}
  \bibinfo{editor}{\bibfnamefont{R.~A.} \bibnamefont{Webb}}
  (\bibinfo{publisher}{Elsevier and Amsterdam}, \bibinfo{year}{1991}), p.
  \bibinfo{pages}{169}.

\bibitem[{\citenamefont{Braun et~al.}(2005)\citenamefont{Braun, Konig, and
  Martinek}}]{braun}
\bibinfo{author}{\bibfnamefont{M.}~\bibnamefont{Braun}},
  \bibinfo{author}{\bibfnamefont{J.}~\bibnamefont{Konig}}, \bibnamefont{and}
  \bibinfo{author}{\bibfnamefont{J.}~\bibnamefont{Martinek}},
  \bibinfo{journal}{Europhys. Lett.} \textbf{\bibinfo{volume}{72}},
  \bibinfo{pages}{294} (\bibinfo{year}{2005}).

\bibitem[{\citenamefont{Ralph et~al.}(1995)\citenamefont{Ralph, Black, and
  Tinkham}}]{ralph}
\bibinfo{author}{\bibfnamefont{D.~C.} \bibnamefont{Ralph}},
  \bibinfo{author}{\bibfnamefont{C.~T.} \bibnamefont{Black}}, \bibnamefont{and}
  \bibinfo{author}{\bibfnamefont{M.}~\bibnamefont{Tinkham}},
  \bibinfo{journal}{Phys. Rev. Lett.} \textbf{\bibinfo{volume}{74}},
  \bibinfo{pages}{3241} (\bibinfo{year}{1995}).

\end{thebibliography}

\end{document}